\pdfoutput=1
\documentclass[%
nofootinbib,
bibnotes,
 amsmath,amssymb,
 aps,
lengthcheck,%
]{revtex4-1}
\usepackage{graphicx}
\usepackage{dcolumn}

\newcommand{\etal}{{\it et al.}}

\begin{document}
\title{Ramsey interferometry using the Zeeman sublevels in a spin-2 Bose gas}
\author{M. Sadgrove$^*$, Y. Eto, S. Sekine, H. Suzuki, and T. Hirano}
\affiliation{Department of Physics, Gakushuin University, Mejiro 1-5-1, Toshima-ku, Tokyo, Japan.}
\email{mark@cpi.uec.ac.jp}

\date{\today}
\begin{abstract}
We perform atom interferometry using the Zeeman sublevels of a spin-2 Bose-Einstein condensate of $^{87}$Rb. The observed fringes are strongly peaked, and 
fringe repetition rates higher than the fundamental Ramsey frequency are found in agreement with a simple theory based on spin rotations.
With a suitable choice of initial states, the interferometer could function as a useful tool for magnetometry and studies of spinor dynamics in general.
\end{abstract}

\maketitle
The study of spinor properties of matter was extended to the field of dilute Bose-gases after the achievement of optical
confinement of a Bose-Einstein condensate (BEC)~\cite{SK1}. This technique gave rise to a number of intriguing studies at the intersection of fields 
as diverse as magnetometry, symmetry breaking and pattern formation~\cite{SK2,SK3,SK4}. More recently, spinor~\cite{TwinAtom} and effective 
spinor~\cite{Gross} condensates (i.e. two mode condensates which can formally be treated as spin 1/2 systems) along with atom-atom interactions 
have provided the tools to create ``non-linear" atom interferometers, that is, interferometers where noise in the quantity of interest can be 
suppressed by using spin-squeezed states~\cite{Ueda,Hyllus}, leading to measurement sensitivity in excess of that classically permitted.

Spin rotations for spin $>1/2$ have been made use of for decades in molecular experiments~\cite{RamseyMolBook}.
Moreover, in terms of interferometry experiments, sharply peaked fringes were observed in an atom 
interferometer using magnetic sublevels in~\cite{Weitz}. In that study the source of the fringe pattern was 
explicitly linked to multi-path interference. Additionally, Ramsey interferometry in a spin-1 atomic gas has
been used in sensitive magnetometry experiments, 
with the aim of detecting a permanent electric dipole moment~\cite{Budker}. Of relevance to our 
particular experiment, we note that rotation and more general manipulation of spin states have been used as 
tools in various studies of spin-2 spinor BECs (see for example ~\cite{SK2,SpinTrans,Skyrmion}). 

Here, we introduce a spinor Bose-gas interferometer using a spin-2 condensate of $~^{87}$Rb, in line with recent 
theoretical proposals~\cite{SpinorBECInt1,SpinorBECInt2}. We observe predicted non-trivial fringe structure, 
with sharp fringes being a particular feature of interest~\cite{SpinorBECInt1}, and, as we will show, fringe oscillations at multiples of the fundamental
period being another. Parallels with proposals for spin echo experiments in spinor BECs can also be made~\cite{Tsubota}. 
We note that our work shares common tools and certain similarities with the experiments mentioned in the above paragraph, but our principle aim
is to demonstrate the possibilities of Ramsey type interferometry in a spin-2 BEC setting. 
We also note that while non-destructive imaging techniques can extract precise information about the atomic precession frequency~\cite{SK2}, 
performing an interferometer sequence followed by a full Stern-Gerlach (SG) measurement as we do here can provide more information about the system. 

\begin{figure}
\centering
\includegraphics[width=0.9\linewidth]{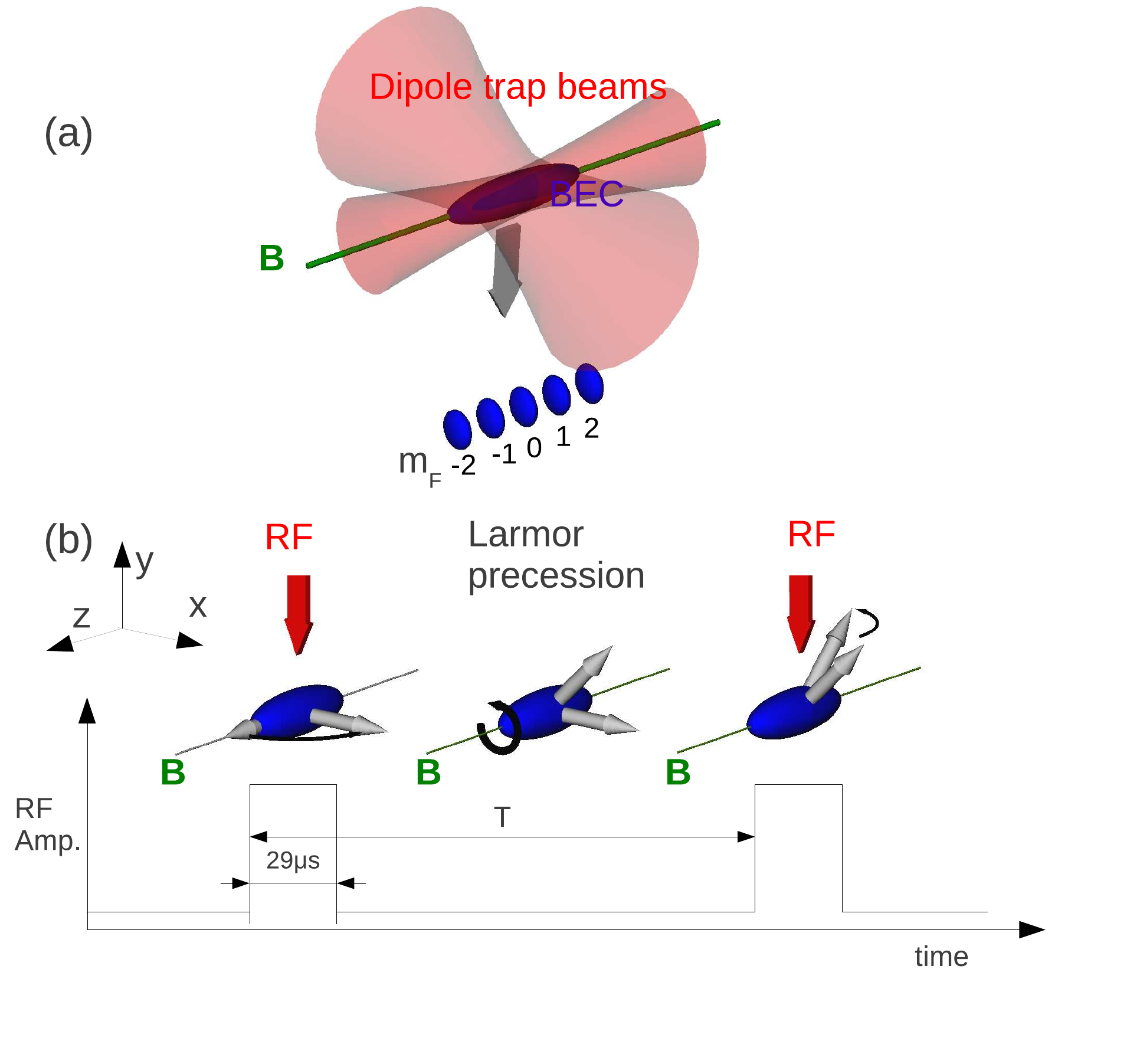}
\caption{\label{fig:exp} Schematic diagram of the experiment. (a) Depicts the BEC confined in a crossed-dipole trap where the interferometer sequence takes place. 
The arrow below the BEC indicates its movement under gravity after it is released from the trap, and its subsequent splitting into five components 
(labelled $m_F = -2,\ldots,2$)
due to an applied gradient magnetic field. (b) shows the interferometer sequence in more detail. Grey arrows depict the atomic spin, while the 
RF field is indicated by a red arrow. Note that the axis of the bias magnetic field is shown by a thin green line labeled ``B" in both (a) and (b).}
\end{figure}

The concept of the experiment is illustrated in Fig.~\ref{fig:exp}. The five magnetic sub-level components associated with a spin-2 gas can be
separately measured, and the population of each component depends on coherent rotations applied by a radio frequency (RF) field and also Larmor precession.
In our experiment, a BEC of $^{87}$Rb is created in a magnetic trap using standard RF evaporation methods.  After loading into a crossed optical dipole trap, typically 4$\times10^5$
atoms remain, and the trapping potential has axial and radial frequencies of order $\sim 33$Hz and $\sim 100$Hz respectively. 
The spin orientation of the optically trapped condensate
is preserved by application of a 20G bias magnetic field (B-field) and the optically trapped BEC's internal state 
is given by $|F=2, m_F=+2\rangle$. The $m_F$ components may be separated using a SG technique as shown in Fig.~\ref{fig:exp}(a).
After a 200ms hold time in the optical trap, we ramp the bias field down to 300mG over 10ms, and apply the interferometer sequence depicted in Fig.~\ref{fig:exp}(b). 
The 300mG field is applied by separate coils, the current supply for which is a commercial laser diode supply with current ripple on the order of 10$\mu$A. We found that a stable bias field 
was very important for achieving repeatable spin rotations.

During the interferometer experiment, an RF $\pi/2$ pulse rotates the atomic 
spin until it is perpendicular to the bias B-field. Subsequently, Larmor precession is allowed to occur for a time $T$. We calculated the Larmor frequency to be 
$f_0 = g_F \mu_B B/h = 210$kHz at our estimated bias B-field of $300$mG, where $g_F$ is the g-factor, and $\mu_B$ is the Bohr magneton. Finally a second RF $\pi/2$ pulse 
resonantly rotates the spin again, and the population of each spin component is read out using the SG technique followed by absorption imaging of the condensate.
Further details of the experimental setup may be found in Refs.~\cite{Kuwamoto,TJ}.

\begin{figure}
\centering
\includegraphics[width=0.9\linewidth]{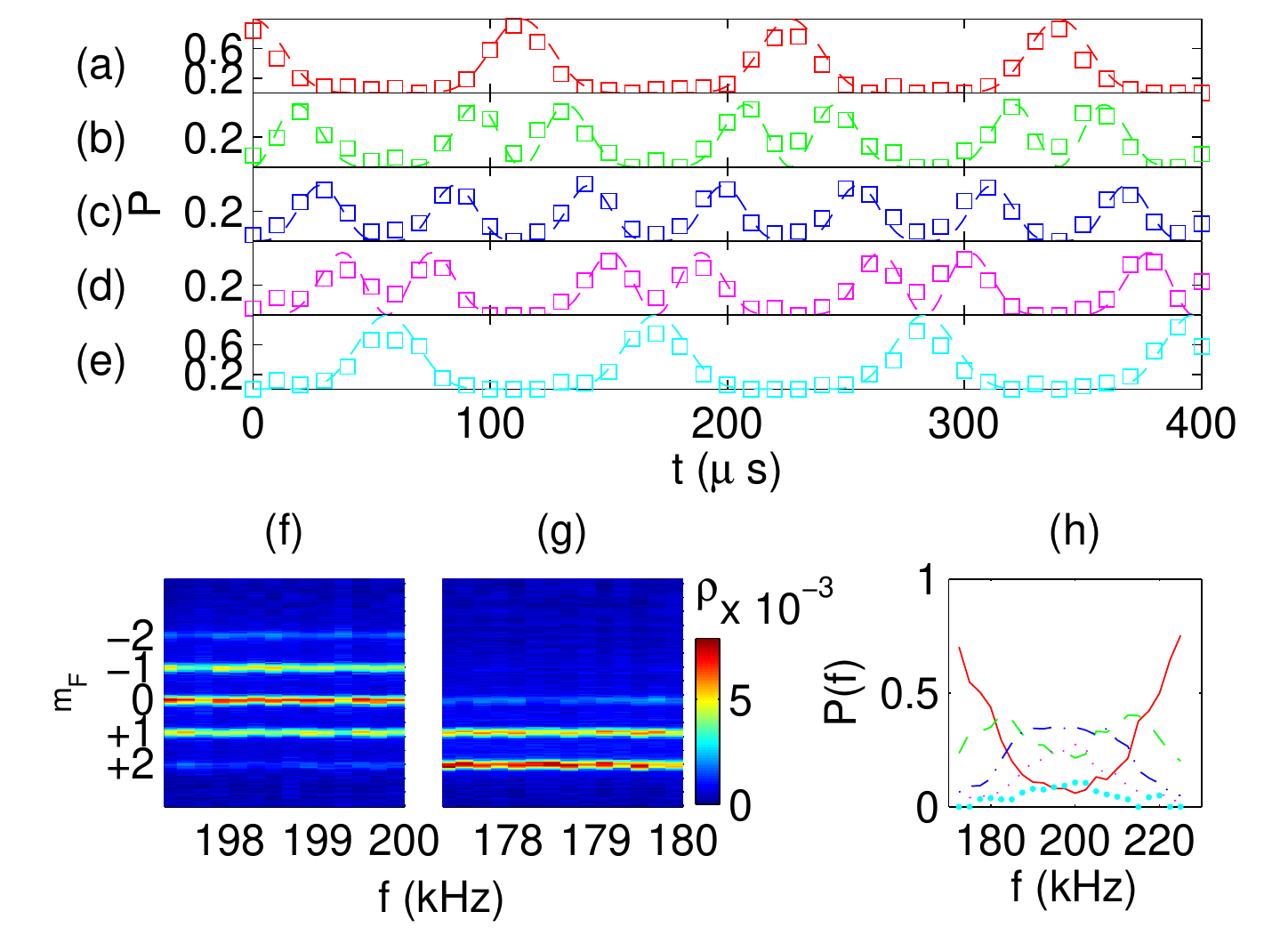}
\caption{\label{fig:rot}Coherent spinor rotation of the BEC. (a) - (e) show the behavior of $P=|\Psi_f|^2$ for the $m_F$ components
+2,-+,0,-1,-2 respectively. In each case squares show the experimental data and dashed lines show the theory of Eq.~\ref{eq:rot}.
(f)  and (g) show the effect of changing the RF frequency on column densities $\rho$ for a nominal $\pi/2$ RF pulse of 29$\mu$s in width for the near resonant and
off-resonant cases
respectively. Finally, (h) shows the component-wise
distribution of $P=|\Psi|^2$ for a detailed scan over the resonant frequency. The populations of the $m_F=+2,+1,0,-1$ and $-2$ states are shown by solid, dashed, dash dotted,
fine dotted and large dotted lines respectively.}
\end{figure}

Before presenting our experimental results, we offer a phenomenological theory of the experiment described above. 
Our experiment realizes a Ramsey-type interferometer over the five Zeeman sublevels of our $^{87}$Rb condensate.
The usual treatment of such systems assumes a two level system and indeed, the theoretical treatments of spinor BEC interferometers so far 
transform into an effective two level system using the Majorana representation~\cite{Majorana, BlochRabi}.

Here, we adopt a different approach, and derive a formula for the interferometer fringe (i.e. population as a function of RF wave frequency) in each sub-level.
We rely on the following two facts: (i) An RF pulse with a frequency near to the Larmor frequency of the atoms induces a coherent rotation of the atoms about an axis perpendicular
to the B-field~\cite{Higbie,Chang,Liu} and (ii) the rotation effected by the RF pulse drops off as a smooth function of the RF detuning from the Larmor 
resonant frequency. 

We tested both of these assumptions experimentally and the results are shown in Fig.~\ref{fig:rot}. In Fig.~\ref{fig:rot}(a-e), measured populations (squares) in $m_F$ sublevels $+2,\ldots,-2$
respectively are compared with theoretical values. The theoretical values may be obtained by recalling that for non-trivial rotations $\beta$ (i.e. rotations which couple $m_F$ sublevels)
the Wigner rotation matrix using the Euler convention is, $D^2(\alpha=0,\beta,\gamma=0)=d^2(\beta)$ where $d^2(\beta)$ is the so-called reduced rotation matrix~\cite{Sakurai}. 
Applying this operation to our initial state (all atoms in the $m_F = +2$ sublevel) $\Psi_i \equiv [\psi_{+2},\psi_{+1},\psi_{0},\psi_{-1},\psi_{-2}]^T=[1,0,0,0,0]^T$ 
and squaring gives the measured probability distribution
\begin{eqnarray}
\label{eq:rot}
|\Psi_{\pi/2}|^2 & = & \left[\frac{1}{16}(1+\cos(\beta))^4, \frac{1}{4}(1+\cos(\beta))^2\sin^2(\beta), \frac{3}{8}\sin^4(\beta),\right.\nonumber \\ 
 & & \left.\frac{1}{4}(1-\cos(\beta))^2\sin^2(\beta), \frac{1}{16}(1-\cos(\beta))^4 \right]^T.
\end{eqnarray}
For comparison with experiment
we take $\beta \equiv \omega t$, where $t$ is the RF pulse width and $\omega$ is the angular rotation frequency of the spin
induced by the RF pulse. As seen in Fig.~\ref{fig:rot}(a-e), the experimental values agree very well with theory for a measured value of $\omega = 2\pi\times8.8$kHz.
We can therefore be confident that the RF pulses perform coherent rotation of the spin as expected, and that the second order Zeeman effect is negligible on the 
time scale considered here. In that case, the five separate fringe patterns produced (one for each Zeeman sub-level $m_F$) 
are non-sinusoidal, and their shape is predicted by the properties of the Wigner rotation matrix $D^2$~\cite{Sakurai}. 
The peak structure can be explained by invoking the spin-2 rotation properties of the BEC, rather than an explicit multi-path interferometer formulation.

Regarding our second assumption, Figs.~\ref{fig:rot}(f-h) show the effect of a nominal $\pi/2$ pulse as the frequency is 
swept over the resonant Larmor frequency $f_0$. The rotational response of the spin
induced by the pulse is seen to tend to zero as the size of the detuning of the RF frequency $f$ from $f_0$ increases. The width $\Delta$ of the frequency response is seen to be 
$\Delta \simeq 25$kHz. By analogy with the standard Ramsey interferometer, we model the frequency dependence by a sinc function. 
In particular,
\begin{equation}
\label{eq:pi2res}
\beta(f) = \frac{\pi}{2}\eta(f) = \frac{\pi}{2}\frac{\Delta\sin[\pi(f-f_0)/\Delta]}{(\pi(f-f_0))}.
\end{equation}
The frequency dependence of the interferometer $\pi/2$ pulses can now be modeled by the operator $d^2[(\pi/2)\eta(f)]$.
\begin{figure*}
\centering
\includegraphics[width=0.9\linewidth]{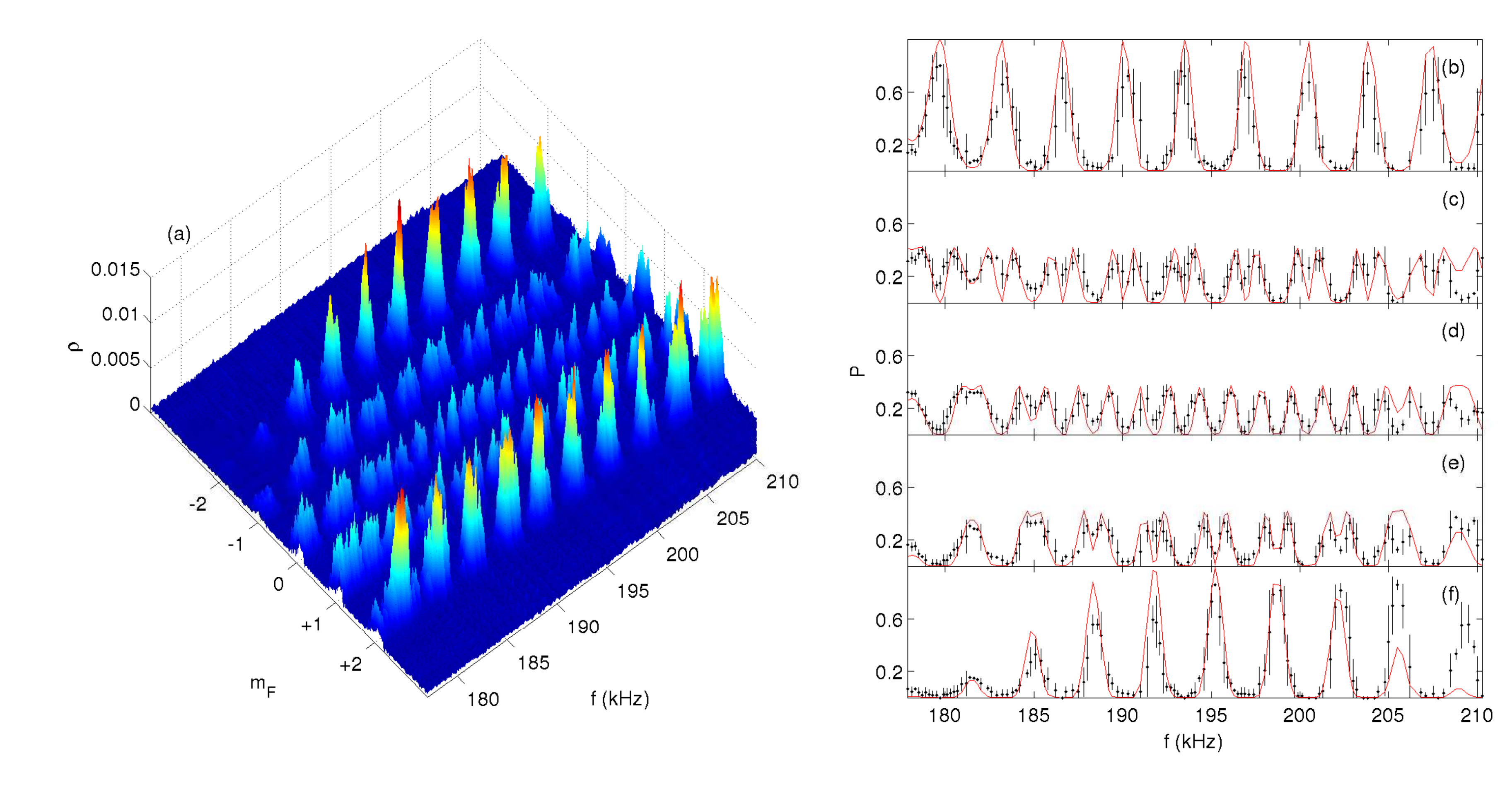}
\caption{\label{fig:int} Measured interference fringes. Interference fringes are shown for a pulse separation $T=285\mu$s.
(a) shows raw column densities $\rho$ from absorption images of the BEC. The population
is seen to oscillate between the $m_F=\pm2$ states as a function of frequency with the amplitude flowing through the $m_F=+1,0$, and $-1$ states. (b-f) 
show relative population 
measurements $P$ for $m_F=+2,\ldots,-2$ respectively, with discrete points showing the experimental results (each being an average over the results at three neighboring frequencies), error bars showing the standard deviation over these three points and solid lines showing the fitted theory from Eq.~\ref{eq:fringe} in each case. 
The fitted parameters are $f_0=195$kHz and $T=290\mu$s and the peaks have a phase offset of $0.14$ radians.}
\end{figure*}
Applying this operator to the initial state $\Psi_i$ models the application of the interferometer's first $\pi/2$ pulse.

To model the entire interferometer, we also need to model the phase accumulation due to Larmor precession between pulses and the phase difference between the 
RF pulses after the interrogation time $T$ has elapsed. Both of these effects can be modeled by rotations about the B-field axis by applying the rotation 
matrix $D^2(\Phi(f,T),\beta\equiv0,\gamma\equiv0)$, where $\Phi(f,T) = 2\pi(f+f_0)T$ is the combined phase accumulation due to Larmor precession and the time between pulses.
Since $\beta=0$, the rotation matrix is diagonal with matrix elements $D^2_{m_F,m_F}=\exp(-{\rm i} m_F\Phi)$.
Finally, 
the rotation operator $d^2[(\pi/2)\eta(f)]$ is applied again to give the final output state. To evaluate the interference fringe, we apply
the Born rule to the complex amplitudes for each $m_F$ component to find the relative population 
$|\Psi_f|^2 = |d^2[(\pi/2)\eta(f)]D^2(\Phi,0,0)d^2[(\pi/2)\eta(f)]\Psi_i|^2$. Performing the matrix multiplication gives
\begin{widetext}
\begin{eqnarray}
\label{eq:fringe}
 |\Psi_f|^2 &= &\left[\left[\cos^2\left[\frac{\pi}{2}\eta\right]\cos^2\left(\frac{\Phi}{2}\right) + \sin^2\left(\frac{\Phi}{2}\right)\right]^4,
4\cos^2\left(\frac{\Phi}{2}\right)\sin^2\left[\frac{\pi}{2}\eta\right]\left[\cos^2\left[\frac{\pi}{2}\eta\right]\cos^2(\frac{\Phi}{2}) + \sin^2\left(\frac{\Phi}{2}\right)\right]^3,\right.\nonumber\\
 & & \frac{3}{8}\left[\cos^4\left(\frac{\Phi}{2}\right)\sin^2\left[\pi\eta\right] + \sin^2\left[\frac{\pi}{2}\eta\right]\sin^2\left(\Phi\right)\right]^2, \nonumber\\
 & & 4\left[\cos^2\left[\frac{\pi}{2}\eta\right]\cos^8\left(\frac{\Phi}{2}\right)\sin^6\left[\frac{\pi}{2}\eta\right] + \cos^6\left(\frac{\Phi}{2}\right)\sin^6\left[\frac{\pi}{2}\eta\right]\sin^2\left(\frac{\Phi}{2}\right)\right],
\left. \cos^8\left[\frac{\Phi}{2}\right]\sin^8\left[\frac{\pi}{2}\eta\right]\right],
\end{eqnarray}
\end{widetext}
where we have left the frequency dependence of $\Phi$ and $\eta$ out of the notation for brevity. For comparison with experiments, we allow an additional free phase parameter
$\phi$ in the expression for $\Phi$. This phase depends linearly on $T$ alone and its effect is 
to shift the position of the peaks without changing their spacing.
\begin{figure}
\centering
\includegraphics[width=0.9\linewidth]{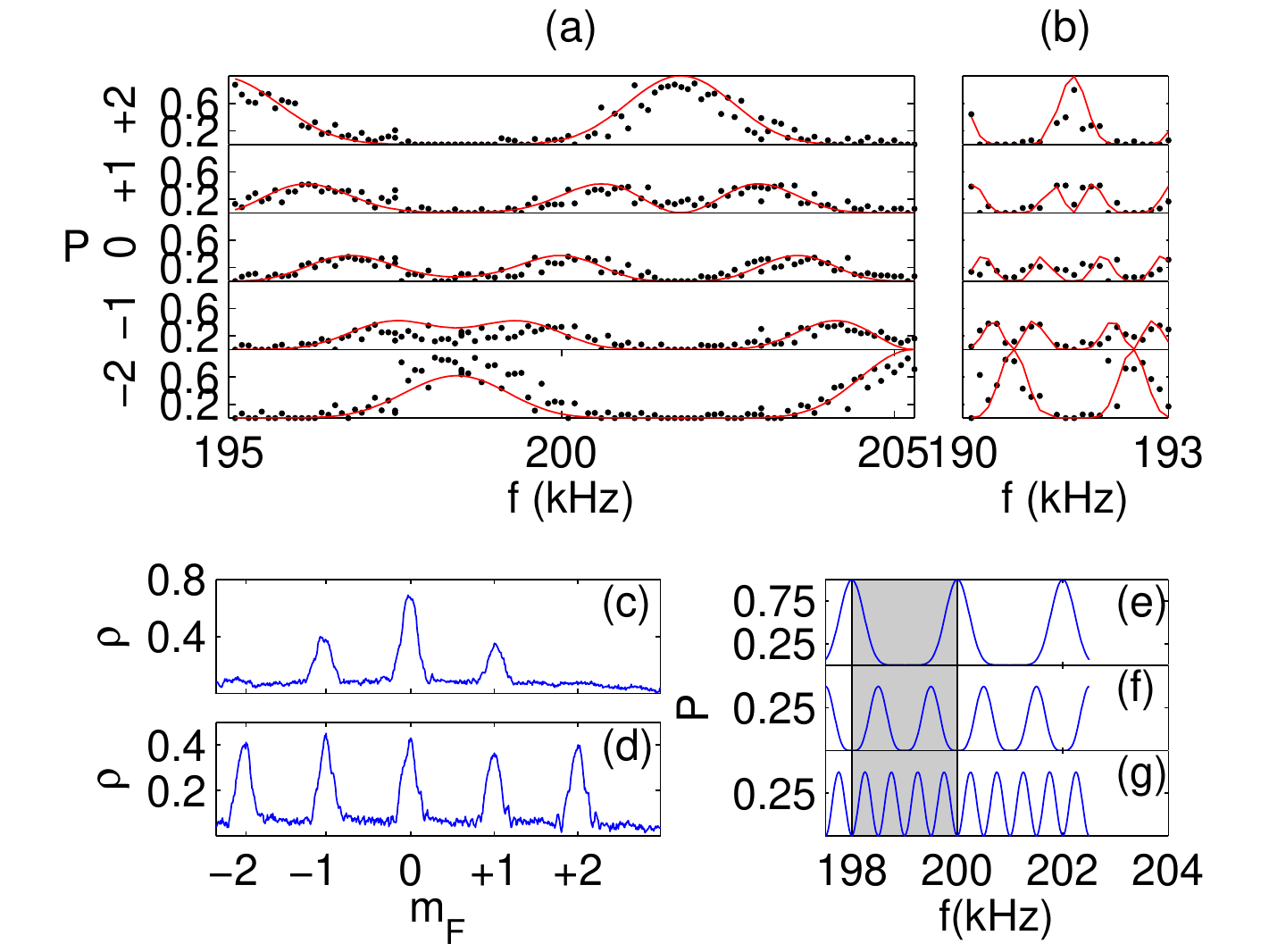}
\caption{\label{fig:atomatom}The effect of interrogation time and input state. Fringes over approximately one period showing the effect of different interrogation time $T$. 
(a) shows the populations for a nominal pulse separation of $T=152\mu$s, 
with $m_F$ as indicated on the vertical axis label. (b) shows data for a nominal pulse separation of $T=530\mu$s. Note that the horizontal scales in each plot are the same for ease of comparison. 
As in Fig.~\ref{fig:int}, discrete points show experimental measurements (this time without averaging) and lines show a fit of Eq.~\ref{eq:fringe} to the data. The fitted parameters 
are $f_0=207$kHz and $T_{\rm fit}=143\mu$s for (a) and $f_0=192$kHz and $T_{\rm fit}=574\mu$s for (b) and the fringes were found to have phase offsets of -0.97 and 2 radians for (a) and (b) respectively.
(c) and (d) show the effect of much longer interrogation times where phase changes due to effects other than rotation cannot be neglected. (c) shows the column density after a single $\pi/2$ pulse and 
5ms hold time in the trap. (d) shows the column density for the same situation as (c) but with a final $\pi/2$ pulse applied at the end of the sequence. Note that the
essentially uniform distribution of amplitude among states cannot be arrived at by simple rotations. Lastly, Figs. (e-g) show how the interferometer can be improved by 
using a different initial state. (e) and (f) show fringes for the $m_F = +2$ and $m_F=0$ states respectively, in the usual case where the initially populated state is $m_F = +2$.
(g) shows the $m_F = 0$ fringe for the case where the initial state is $|m_F=+1\rangle$. In this case, we see that 4 times as many fringes per period are expected compared 
with the fundamental periodicity of the interferometer. The gray shaded region shows one period (i.e. $2\pi$ phase change) in each case.}
\end{figure}

We now move on to our experimental interferometry results.
To perform interferometry on the BEC, we fixed a value $T$ for the time between the RF pulses (the interferometer \textit{interrogation time}) and observed the relative population of the Zeeman sublevels 
as the pulse frequency $f$ was varied between 175 and 210kHz. Fig.~\ref{fig:int}(a) shows normalized column densities calculated from absorption images of the BEC for a range of frequencies. From this essentially raw data, the oscillatory behavior is clearly visible, but the detailed structure is not readily apparent.
To see the fringe structure, we calculated the relative population in each substate (shown as discrete points in Figs.~\ref{fig:int}(b-f)).
We then fitted Eq.~\ref{eq:fringe} to the experimental result for $m_F= +2$ ( which has the best signal to noise ratio) and
used the obtained parameters in Eq.~\ref{eq:fringe} for all $m_F$ values. The fits are shown by solid lines in Figs.~\ref{fig:int}(b-f). The fitted parameters agree with the experimentally measured 
parameters to within $3\%$ with a small extra phase shift of $\phi=0.14$ radians being necessary to account for the exact peak position.

The experimentally measured fringe pattern is non-sinusoidal and in good qualitative agreement with the predictions of Eq.~\ref{eq:fringe}. It is interesting to note
that the $m_F = +1,0$ and $-1$ component fringes contain oscillations with a smaller period than the fundamental seen in $m_F  = \pm2$. In particular, the $m_F=0$
fringe oscillates with essentially half the period of the principle oscillation. This kind of behavior is also found in multi-beam optical interferometers~\cite{ThreeBeam}
and is a logical consequence of the fact that each peak in the population of $m_F = \pm2$ is defined by two zero-crossings (i.e. as the amplitude crosses through the 
intermediate $m_F$ states).

We estimated the sensitivity of the interferometer by finding the standard deviation over three nearest neighbor frequencies for every second point in our frequency scan.
Then, for each peak in the highest visibility $m_F=+2$ fringe pattern, we found the smallest distinguishable phase difference within the experimental error. 
In this way, we calculated the average phase sensitivity to be about 0.6 radians or $10\%$ of a fringe. 
Given our relatively short interrogation time, this result is comparable with other Ramsey atom-interferometers
operating on coherent states~\cite{Altin}. We would like to note, however, that utilizing all five fringes of our spin-2 system presents an advantage relative to
spin $1/2$ or spin $1$ systems because at any given frequency five fringes are available. By choosing the fringe with the steepest rate of change at the desired 
frequency, sensitivity can be maximized. Additionally, we note that engineering the initial distribution of $m_F$ states can produce high visibilities in 
states other than $m_F=\pm2$.

We also tested the interferometer for shorter and longer interrogation times, as seen in Figs.~\ref{fig:atomatom}(a) and (b). The fringes show typical Ramsey-type behavior
with the period becoming longer or shorter as the interrogation time was decreased or increased respectively. Details are given in the caption of Fig.~\ref{fig:atomatom}.
We found that at a larger interrogation time 
(5ms) the fringe repetition rate becomes very high making accurate measurements of the fringe pattern difficult. Indeed, we found no clear pattern for the $T=5$ms
measurements, and the exact pattern for a given RF frequency was not perfectly repeatable. However, we note that \emph{qualitatively} the distributions output by the interferometer
at longer interrogation times are very different from the results for
$T$ of order 100$\mu$s. Figs.~\ref{fig:atomatom}(c) and (d) show column densities for a 5ms interrogation time in the case of a single $\pi/2$ pulse and with 
the inclusion of the second $\pi/2$ pulse respectively. As Fig.~\ref{fig:atomatom}(c) shows, any higher order processes do not make themselves known 
in the distribution shape after $5$ms, and indeed we have found that it takes almost an order of magnitude longer before $m_F$ changing collisions produce 
significant population in the $m_F=\pm2$ states.

The result of applying a second $\pi/2$ pulse after 5ms is shown in Fig.~\ref{fig:atomatom}(d). The column density distribution in this typical case
shows an essentially uniformly distributed probability over the $m_F$ states. Such a distribution is impossible to produce with coherent rotations alone.
Dephasing due to the non-zero magnetic field gradient in the experiment is the likely cause of this behavior.
We note in passing that the uniform distribution in Fig.~\ref{fig:atomatom}(d) is reminiscent of spin transport results seen in~\cite{SpinTrans},
where equipartition over spin sub-levels was found after a sufficiently long time. 

We now comment on the possibility of using different initial states as inputs to the interferometer. 
As seen in Figs.~\ref{fig:atomatom}(e-g), simulations show that merely changing the initially populated input port of the interferometer should produce non-trivial behavior 
in the fringes. In particular, taking $|\Psi_i\rangle = |m_F=+1\rangle$ produces fringes in the $m_F=0$ sublevel with \emph{four times} the principle repetition rate of the interferometer. We have already 
measured the doubling of the repetition rate when the initial state is $|m_F=+2\rangle$, but the usefulness of this effect is debatable since the amplitude of the fringes 
is also halved in the $m_F=0$ case. However, a quadrupling of the fringe repetition rate with no further loss in amplitude, as predicted here, should be useful since it provides 
four times the number of fringes to measure as compared to the usual case. 

In order to overcome the quantum limit of sensitivity, the current interferometer would in principle need
to use squeezed input states. The relative ease of using techniques such as those of~\cite{TwinAtom} to produce number-squeezed input states 
in our setup is promising, although calculations are needed to confirm the precise benefits of such ``twin atom" states for our interferometer.
We would like to note, however, that even without considering exotic input states, improvements to sensitivity can be made, as seen in the predicted increase
in the fringe repetition rate when the input state is in the $m_F = +1$ magnetic sublevel (see the previous paragraph).
Such increased rapidity of fringe oscillation compared with the fundamental rotation 
frequency is often associated with entangled state interferometry~\cite{Walther}, but here arises naturally due to the multi-port nature of our interferometer.
One motivation for increasing the sensitivity is magnetometry: Because $f_0$ depends on the local magnetic field,
changes in $B$ could be detected by looking for interferometric fringe shifts. 

We emphasize that the present results 
serve mainly to establish the possibility of high visibility Ramsey interferometry in a spin-2 Bose gas.
Although the observed sensitivity is in line with results from other interferometry experiments, technical sources of noise need to be eliminated before
the device can be used for precision measurements. For interrogation times of order 1ms, dephasing effects come into play. 
In particular, the current experiment is limited by the non-zero magnetic field gradient in the vacuum cell ($\sim 30$mG/cm), and time-dependent 
fluctuations in the ambient magnetic field (amplitude $\sim100\mu$G). Such effects can be partially corrected for by using a spin echo $\pi$-pulse between the two Ramsey pulses used in 
the current experiment. Indeed, the overall goal for the present experiment is to suppress sources of technical noise, and use such spin-echo pulse sequences for detection
of AC magnetic fields~\cite{Lukin}. As we will report elsewhere, such noise suppression and the application of more sophisticated pulse sequences can lengthen the interrogation time to several ms
~\cite{Eto}. 

In conclusion, the atom interferometer realized here by applying coherent rotations to a spin-2 Bose gas, essentially functions as a multi-port interferometer with non-trivial
fringes seen at each port. The measured fringe patterns demonstrate properties in good agreement with a phenomenological theory based on spinor rotation.
The measured sensitivity is reasonable and may be improved firstly by eliminating technical noise in the imaging of the BEC, and ultimately
by engineering the input state of the interferometer. 
Overall, our results provide a useful new tool for probing the spin coherent properties of Bose-Einstein condensates. 

We acknowledge support from the Program for World-Leading Innovation Research and Development on Science and Technology (FIRST),
and Grants-in-Aid for Young Scientists (B) (No. 23740314) from the Ministry
of Education, Culture, Sports, Science, and Technology of Japan.

\end{document}